# Analytical expression for Risken-Nummedal-Graham-Haken instability threshold in quantum cascade lasers

**N. VUKOVIC,**[1,3] **J. RADOVANOVIC,**[1] **V. MILANOVIC,**[1] **AND D. L. BOIKO**[2,4]

[1] *School of Electrical Engineering, University of Belgrade, Bulevar kralja Aleksandra 73, 11120 Belgrade, Serbia*
[2] *Centre Suisse d'Électronique et de Microtechnique (CSEM), Jaquet-Droz 1, CH-2002 Neuchâtel, Switzerland*
[3] *nikolavukovic89@gmail.com*
[4] *dmitri.boiko@csem.ch*

**Abstract:** We have obtained a closed-form expression for the threshold of Risken-Nummedal-Graham-Haken (RNGH) multimode instability in a Fabry-Pérot (FP) cavity quantum cascade laser (QCL). This simple analytical expression is a versatile tool that can easily be applied in practical situations which require analysis of QCL dynamic behavior and estimation of its second threshold. Our model for a FP cavity laser accounts for the carrier coherence grating and carrier population grating as well as their relaxation due to carrier diffusion. In the model, the RNGH instability threshold is analyzed using a second-order bi-orthogonal perturbation theory and we confirm our analytical solution by a comparison with the numerical simulations. In particular, the model predicts a low second threshold in QCLs. This agrees very well with experimental data available in the literature.

## References and Links

## 1. Introduction

Nowadays quantum cascade lasers (QCLs) emitting in the mid-infrared (MIR) spectral range find widespread applications in health care, gas sensing, telecommunications and broadband spectroscopy [1]. While many applications require QCLs operating strictly in a single mode regime, QCLs producing frequency combs are proven to be very efficient in spectroscopic applications [2]. However, the majority of QCL based frequency combs do not exhibit phase locking relationship between individual modes. So far, a very moderate progress was made towards achieving ultra-short pulse production regimes in QCLs operating under room temperature conditions [3]. At the same time, ultra-short optical pulse generation by QCLs would enable applications in other important domains such as time resolved pump-probe MIR spectroscopy of rotovibrational modes of molecules or the environmental remote sensing of molecules in the mid-infrared spectral range.

One very interesting proposal on how to produce short MIR pulses in a QCL [4] has emerged when a multimode Risken-Nummedal-Graham-Haken (RNGH) like instability [5,6] (or so-called "second threshold" of a laser) was experimentally observed in these lasers. Although the interpretation of measured autocorrelation traces in particular QCLs was not straightforward, one can expect that the lasing dynamics in a QCL might be tailored so as to produce regular RNGH self-pulsations.

Interestingly, low-threshold RNGH instability was observed both in ridge waveguide QCLs and in buried heterostructure QCLs [4,7,8]. These experimental observations strongly disagree with the second threshold predicted by the RNGH theory [5,6]. Although the role of spatial hole burning (SHB) in lowering the second threshold was recognized, a conclusion was drawn in [7] that RNGH instability does not occur just as a result of the induced grating of carrier population. In [7], an additional assumption was made about a built-in saturable absorber in the cavity of QCL, which reduces its second threshold from the 9-fold excess as predicted in [5,6] to about 1.1 times above the lasing threshold as it was experimentally measured. As a matter of fact, the saturable absorber has been shown to lower the instability threshold in a laser [9]. However, the nature of saturable absorption in QCLs has never been fully clarified [1]. In particular, in [7], the Kerr lensing effect has been proposed as a possible mechanism responsible for the saturable absorption. In case of narrow ridge waveguide lasers, due to the overlap of the waveguide mode tails and the metal contact deposited on the waveguide, the Kerr lensing may lead to a saturable absorption. However it cannot produce the saturable absorption effect of the same strength in buried heterostructure QCLs. At the same time low-threshold RNGH instability was observed in both ridge waveguide and buried heterostructure QCLs.

In our recent study [10] we have analyzed conditions for multimode RNGH instability in a Fabry-Pérot cavity laser and found that a set of inaccuracies was admitted in the previous treatment [7]. As a result we propose an alternative mechanism responsible for low second threshold in QCLs. It also favors the second threshold lowering by the SHB effect but does not require a saturable absorber. More specifically we show that a combined effect of the carrier coherence grating, carrier population grating and relaxation processes due to carrier diffusion leads to the onset of multimode instabilities at Rabi flopping frequency. We consider such spectral behavior as an evidence of RNGH instability. Our approach to calculate the second threshold has shown a convincing agreement with the available literature data and with the results of numerical simulations. From a practical point of view, this can be converted into a useful engineering tool enabling to examine particular QCL design and predict its dynamic behavior or to tailor QCL design for a specific application. In this paper we obtain a simple closed-form analytical expression for the second threshold in a QCL with monolithic FP cavity. The main result of this paper is summarized by Eq. (14). Technically, it is obtained by using the second-order bi-orthogonal perturbation theory applied to the Lyapunov stability analysis of the single-mode lasing regime in a QCL.

## 2. Theory of the 2$^{nd}$ threshold

The RNGH instability in a continuous wave (CW) single-mode laser arises due to Rabi splitting of the lasing transition induced by the lasing mode [11]. As a result of such spectral broadening and reshaping of the gain curve, the laser can provide sufficient optical gain for other longitudinal modes. If below the RNGH instability threshold, the optical mode is fully controlled by the laser cavity, at above the second threshold, the dynamic behavior is very different. The multimode RNGH instability is related to excitation of rapid coupled oscillations of the medium polarization $P$ and population inversion $N$ [10]. The buildup of Rabi oscillations at above the second threshold indicates that now a macroscopic polarization (coherence) has the major impact on the optical field in the laser cavity [6]. The medium polarization thus exhibits the opposite behavior to that which is known in the literature as the adiabatic following approximation [12].

RNGH instability is thus a multi-longitudinal-mode amplitude instability of a laser [13] which occurs on a very fast time scale. It cardinally differs from other possible instabilities in semiconductor lasers such as transversal mode instability in broad area lasers [14] or relatively slow phase instabilities in semiconductor laser [15,16], which admit adiabatic approximation for the medium polarization.

In order to find the gain curve for instabilities in initially non-lasing cavity modes, one applies the linear stability analysis to the Maxwell-Bloch equations (MBE) describing a single-mode CW laser [5,6]. In Appendix A we briefly present the original model equations, while further elaboration is given in [10]. Here we provide a brief account of these. Starting from MBE and accounting for the standing wave pattern of the optical field in the cavity, we first get a truncated system of coupled-mode equations for the field, carriers and medium polarization as well as their spatial harmonics. We then find a stationary solution for the single-mode CW lasing regime and perform its linear stability analysis with respect to small perturbations to the lasing mode. Then using the ansatz from [5], we recast the perturbations in the form of propagating wavelets $\propto \exp(in_g\Omega z/c + \Lambda t)$, where $\Lambda$ can be interpreted as Lyapunov exponent, $\Omega n_g/c$ is the detuning of the propagation constant from the main lasing mode and $n_g$ is the group velocity index (for simplicity, we call the parameter $\Omega$ as a frequency offset [10]). The resulting linearized system of differential equations is defined by a 9×9 block-diagonal matrix $M_{9\times 9}$ [10]. The dispersion of the instability increment Re($\Lambda$) as a function of the detuning $\Omega$ then follows from the eigenvalue problem of $M_{9\times 9}$:

$$\det(M_{9\times 9}(\Omega) - \Lambda I) = 0 \tag{1}$$

where $I$ is the identity matrix. It appears that up to 40-fold excess of the pump rate above the lasing threshold of QCL, only one eigenvalue in Eq. (1) may have a positive real part Re($\Lambda$) [10].

The purpose of this paper is thus to find an analytic expression for the second threshold condition when Re(Λ)>0 and leads to a multimode RNGH instability. Note that the matrix $M_{9\times 9}$ is composed of two main diagonal blocks of 4×4 and 5×5 sizes and the eigenvalue with a positive real part is related to the smaller size (4×4) block, which we show below in Eq. (2). (The remaining 5×5 block exhibits only stable solutions in the range of pump rates that are used in practice [10].) For the reason we explain later, we split this 4×4 matrix into two matrices $M^{(0)}$ and $M^{(1)}$:

$$\|M\| = \|M^{(0)}\| + \|M^{(1)}\| = \begin{bmatrix} -\frac{1}{T_{2,eff}} & -\frac{(2\nu_0-1)}{2T_{2,eff}} & 0 & 0 \\ -\frac{c}{n_g}l_0 & -\frac{c}{2n_g}l_0 - i\Omega & 0 & 0 \\ 0 & 0 & -\frac{1}{T_g} & 2E \\ 0 & 0 & -\frac{E}{2} & -\frac{1}{T_{2,g}} \end{bmatrix} + \begin{bmatrix} 0 & 0 & \frac{E}{2} & 0 \\ 0 & 0 & 0 & 0 \\ -2E & E\left[1+(\nu_0-1)\frac{T_{2,g}}{T_{2,eff}}\right] & 0 & 0 \\ 0 & -\frac{(\nu_0-1)}{2T_{2,eff}} & 0 & 0 \end{bmatrix} \quad (2)$$

where we omit the subscript "4×4" because all matrices that will be discussed from now on will be of the 4×4 size. In Eq. (2), $l_0$ is the cavity loss coefficient that comprises the intrinsic material losses $\alpha_i$ and the output coupling losses; $T_1$ and $T_2$ are the longitudinal and transverse relaxation times of the carriers. As mentioned above, some of the system variables in Eq. (2) are related to the complex amplitudes of the induced spatial gratings such as the carrier coherence grating and the carrier population grating. The overall relaxation rates for these variables account for the relaxation due to carrier diffusion, yielding the effective relaxation times of $T_{2,g} = \left(T_2^{-1} + 9Dk^2\right)^{-1}$ and $T_g = \left(T_1^{-1} + 4Dk^2\right)^{-1}$ [10], where $D$ is the diffusion coefficient for electrons in the plane of optically active QWs and $k=2\pi n_g/\lambda$ is the wavenumber of the main lasing mode in the cavity. In (2), the carrier diffusion slightly speeds up the relaxation of the medium polarization so as the effective transverse relaxation time reads $T_{2,eff} = \left(T_2^{-1} + Dk^2\right)^{-1}$. The other matrix elements in (2) are defined by normalized field amplitude of the optical wave in the cavity

$$E = \sqrt{\frac{p-\nu_0(p)}{2T_1T_{2,eff}}}, \quad (3)$$

The field amplitude in Eq. (3) is normalized in such a way that it represents the Rabi oscillations frequency

$$\Omega_{Rabi} = \sqrt{2}E = \sqrt{\frac{p-\nu_0}{T_1T_{2,eff}}} \quad (4)$$

where the coefficient √2 appears due to the standing wave pattern of the optical field in the cavity. Here $p$ is the pump rate normalized to the lasing threshold in the absence of SHB, while parameter $\nu_0(p)$ accounts for the SHB effect, see also [10]. It increases the (effective) lasing threshold and reduces the slope efficiency:

$$\nu_0(p) = \frac{1}{2}\left(p+1+\frac{T_{2,eff}}{T_{2,g}}+\frac{2T_1T_{2,eff}}{T_gT_{2,g}}\right) - \sqrt{\frac{1}{4}\left(p+1+\frac{T_{2,eff}}{T_{2,g}}+\frac{2T_1T_{2,eff}}{T_gT_{2,g}}\right)^2 - p\cdot\left(1+\frac{T_{2,eff}}{T_{2,g}}\right) - \frac{2T_1T_{2,eff}}{T_gT_{2,g}}} \quad (5)$$

Note that $\nu_0$ is independent of the photon lifetime in the cavity. Figure 1(a) displays the effective threshold behavior in QCLs emitting at 10 and 8 μm (QCL 1 and QCL 2) as compared to a typical quantum well laser diode (QW LD) operating at 820 nm wavelength (green curve). For all considered lasers we assume a simple single-section Fabry-Perot (FP) cavity design and cavity length of 4 mm. In Table 1 we summarize the model parameters [7,17] for the three types of lasers considered here.

In the case of LD, the influence of the SHB effect on the effective lasing threshold Eq. (3) is negligible in the considered range of pump rates (p<3). This can be attributed to the gain recovery time $T_1$ in LDs being three orders of magnitude higher than $T_2$. Our model curves for the effective threshold in QCLs plotted in Fig. 1(a) take into account large dispersion in the relaxation time constants due to possible design variations in the active region of QCLs. In both considered QCL examples, the effect of SHB on the effective lasing threshold is significant. As expected, at the lasing threshold ($p=1$), when the optical field is weak, the SHB does not affects the output characteristics of QCL and $v_0=1$. However at higher pump rates ($p>1$), the SHB effect sets in, yielding an increase of the effective threshold $v_0$.

Table 1. Dynamic model parameters for QCLs and QW LDs considered in this paper

| Parameter | Name | QCL 1 [17] | QCL 2 [7] | QW LD [17] |
|---|---|---|---|---|
| $\lambda$ | Lasing wavelength | 10 μm | 8 μm | 850 nm |
| $T_1$ | Carrier lifetime | 1.3 ps | 0.5 ps | 1 ns |
| $T_2$ | Carrier dephasing time | 140 fs | 67 fs | 100 fs |
| $T_{2,eff}$ | Effective carrier dephasing time in the presence of diffusion | 138 fs | 66 fs | 89 fs |
| $T_g$ | Relaxation time of the carrier population grating | 0.927 ps | 0.403 ps | 0.21 ps |
| $T_{2,g}$ | Relaxation time of the coherence grating | 128 fs | 62 fs | 48 fs |
| $\alpha_i$ | Intrinsic material loss | 24 cm$^{-1}$ | 24 cm$^{-1}$ | 5 cm$^{-1}$ |
| D | Diffusion coefficient | 180 cm$^2$/s | 180 cm$^2$/s | 20 cm$^2$/s (ambipolar) |
| $n_g$ | Group refractive index | 3.3 | 3.3 | 3.8 |
| $R_1, R_2$ | Cavity facet reflection coefficients | 27% | 27% | 27% |
| $\Gamma$ | Optical mode confinement factor | 0.5 | 0.5 | 0.01 |
| $\partial g/\partial n$ | Differential material gain | 2.1×10$^{-4}$ cm$^3$/s | 2.1×10$^{-4}$ cm$^3$/s | 1×10$^{-6}$ cm$^3$/s |
| $n_t$ | Transparency carrier density | 7×10$^{14}$ cm$^{-3}$ | 7×10$^{14}$ cm$^{-3}$ | 2×10$^{18}$ cm$^{-3}$ |

At very high pump rates ($p \gg 1$, not shown in the Fig. 1(a)), $v_0$ asymptotically approaches the value of $1+T_{2,eff}/T_{2,g}$. Note that despite the fact that longitudinal and transversal relaxation times in the two QCL examples differ by a factor of two, the difference in $v_0$ behavior is marginal.

Figure 1(b) illustrates the effect of SHB on the Rabi oscillation frequencies (Eq. (4)) linked to the amplitudes of the optical fields in the cavities of considered LD and two QCL examples. The Rabi oscillation frequencies are plotted with (solid curves) and without (dashed curves) accounting for the SHB effect. In both QCLs, the SHB reduces the Rabi flopping frequency due to the effect on the lasing threshold $v_0$. Thus at the normalized pump rate $p=3$, the reduction in Rabi frequencies of QCLs is about 10% (compare the solid and dashed curves in Fig. 1(b)). However as a consequence of $v_0$ behavior in LD depicted in Fig 1(a), the reduction of the Rabi frequency in a LD is small (red solid and dashed curves overlap in the figure).

In Figure 2(a), on example of a short-cavity QCL, we illustrate the spectral behavior of the instability increment Re($\Lambda$) obtained as an eigenvalue of the linear stability matrix (2). In this example, we use parameters of QCL 1 from Table 1 and we assume that the cavity length is 100 μm. The choice in favor of a shot-cavity QCL is made because of the large spectral separation of the cavity modes, which is of 450 GHz in the considered example. In the Fig. 2(a), the instability increment is multiplied by the cavity round-trip time, yielding thus a round-trip gain of 2Re($\Lambda$)$n_g L/c$ for unstable cavity modes. The gain curves for instabilities in non-lasing cavity modes are depicted at several pump rates ranging from $p$=1.1 to $p$=3.2 times above the lasing threshold. Note that the linear stability matrix Eq. (2) does not account for the dispersion of the cavity and for the round-trip self-repetition condition of the cavity modes. Therefore we also indicate frequencies of the cavity modes by plotting the Airy function of the "cold cavity"

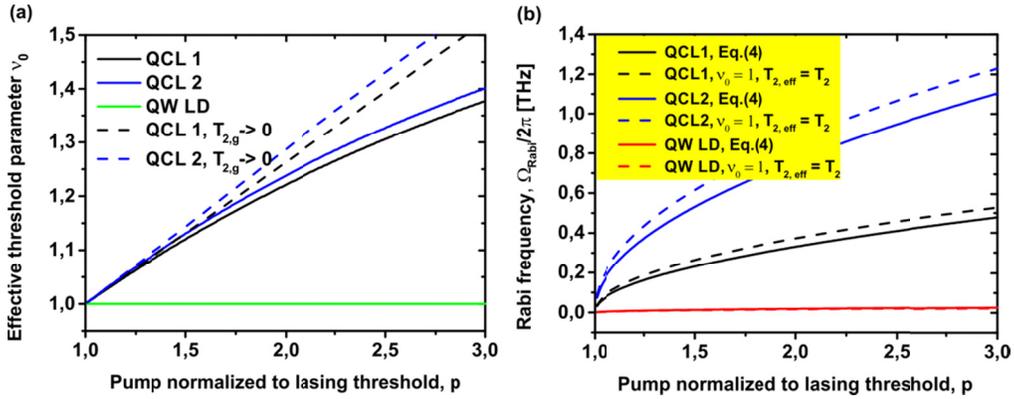

Fig. 1. (a) Threshold parameter $v_0(p)$ plotted as a function of the pump normalized to the lasing threshold for two different sets of QCL parameters (solid black and blue curves) and for a laser diode (green curve). Dashed curves depict $v_0(p)$ behavior without the coherence grating ($T_{2,g} \to 0$, see Eq. (5') in Section 3.2). (b) Their corresponding Rabi frequencies are calculated from Eq. (4) (solid curves) and with $v_0(p)$=1 (dashed curves). The cavity length is 4 mm in all considered examples. All other parameters are listed in Table 1.

(black curve, not in scale). In the Fig. 2(a), the offset frequency $\Omega$ is normalized to the cavity mode separation. The zero offset frequency ($\Omega$=0) corresponds to the initial lasing mode. With increasing pump rate, the instability gain increases and the maximum of the gain shifts to higher frequency. In fact for a given $p$, the spectral maximum of the gain for multimode instability is located at the offset frequency $\Omega_{max} \approx \Omega_{Rabi}$ (further details age given below in the discussion of Fig. 4). We consider such spectral behavior as an evidence of RNGH instability. The original RNGH theory [5,6], which was established for a CW *unidirectional* ring laser, requires that the instability gain is positive for some of the non-lasing cavity modes. In a ring laser, such situation may occur if the pump rate is at least 9 times above the lasing threshold. In FP-cavity QCL, the carrier population grating and the carrier coherence grating reduces the second threshold. As seen from Fig. 2(a), the gain for instability of the first two nearest-neighbor modes is positive already at $p$=1.1. However from the numerical simulations we find that RNGH self-pulsations do not occur at such pump rate.

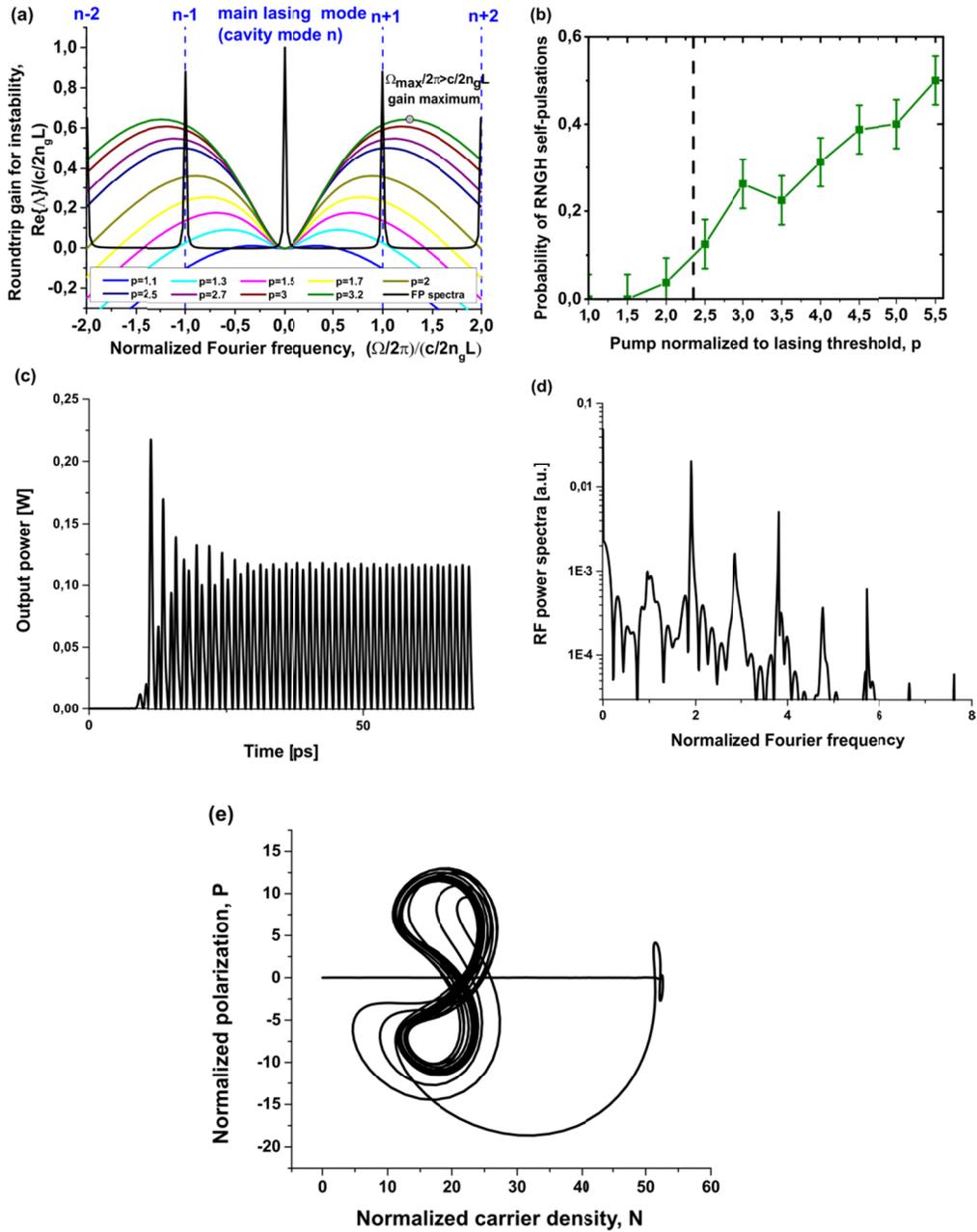

Fig. 2.(a) Round-trip gain spectra responsible for multimode Risken-Numedal-Graham-Haken instability in a short-cavity QCL ($L$=100 μm) are plotted for different pump rates. The zero offset frequency corresponds to the initially lasing cavity mode (the mode with index $n$) when the laser still operates in the single-mode regime. The cavity mode separation is $c/2n_gL$ =0.45 THz (b) The probability of occurrence for RNGH instability is plotted as a function of the pump normalized to lasing threshold $p$. Each data point is based on 80 realizations and therefore the uncertainty is of ±0.056 (vertical bars). Dashed vertical line indicates the second threshold obtained from Lyapunov stability analysis. Example of numerical simulations: (c) Output power waveform, (d) corresponding RF power spectra and (e) P-N attractor for one of the realizations at $p$=3. The parameters used in these simulations are listed in the column labeled "QCL 1" in Table 1.

Figure 2(b) summarizes the results of numerical simulations for QCL depicted in Fig. 2(a). We use the travelling wave (TW) rate equation model that is similar to the one in Ref. [18] but does not assume a saturable absorber in the cavity and accounts for the diffusion of carriers and coherences [10]. Time domain simulations are performed for the QCL being initially in non –lasing state. The TW model incorporates Langevin force terms that seed the spontaneous polarization noise into the system. We vary polarization noise power over six orders of magnitude, from eight to two orders below the polarization noise level in homogeneously broadened ensemble of two-level quantum oscillators (see Eq. (5.5.22) in [19]). In all cases, we have not seen the impact of the noise power on the lasing regime. Only the delay time to the onset of emission was changing. For each pump rate $p$, we perform a series of numerical simulations [10] and count the number of occurrences of RNGH self-pulsations as opposed to the CW lasing regime (for details see Figs. 4. and 5. in [10]). The criterion to determine whether the particular simulation evolves into RNGH self-pulsations is the observation of oscillations in the output power (Fig. 2(c)) and RF power spectrum (Fig. 2 (d)) as well as the peculiar shape of the system attractor in the P-N plane (polarization-carrier density variables) reaching high coherence values P on the system trajectory, that is with the maximum of P approaching a half maximum of N. The P and N variables used in Fig. 2 (e) are introduced following the approach of [18]. In particular, the variable P measures the order parameter in the system (the carrier density in the coherent state) and has the same units as the carrier density N. In Fig. 2(e), both variables P and N are normalized on the transparency carrier density. More details can be found in Ref. [10].

The self-pulsations in a short-cavity QCL are observed at frequencies very close to the cavity round-trip frequency (or its harmonic) and when it (or its harmonic) approaches $\Omega_{Rabi}$ [10]. In Fig 2(b) we show the average probabilities to obtain multimode RNGH self-pulsations. Each point is the result of 80 realizations so as the uncertainty is of ±0.056 (shown as error bars). Note that out of 80 realizations at $p$=1.5, no occurrence is detected with RNGH self-pulsations. The RNGH instability starts to develop at $p$ between 2 and 2.5, with $p$=2.5 being the first point for which such probability exceeds the uncertainty range. We clearly see that this is a higher pump rate than one would expect from the approach of [5,6] based on the positive instability increment for single mode CW lasing [the gain curve at $p$=1.1 in Fig. 2(a)].

The positive increment for instability of single mode CW lasing regime in Eq. (1) indicates the instability of the lasing mode but it does not imply that RNGH self-pulsations will occur. The laser may just switch to another mode, which is also unstable. The increment for instability of the mode is small [e.g. see the cyan curve in Fig.2 (a) obtained at p=1.3] so the switching process will be slow and may be driven by technical noise. A classic example is a ring laser with reciprocal cavity for counter-propagating modes and absence of mode coupling via backscattering in the cavity: it shows sporadic switching of the lasing direction (spontaneous mode-jumps) at times much longer than the characteristic times $T_1$, $T_2$ and $\tau$ [20]. A FP cavity laser may also exhibit such spontaneous mode-jumps [16] and the linear stability analysis in Eq. (1) does not exclude such possibility. Other possibility is related to multimode phase instability in semiconductor laser which admit adiabatic approximation for the medium polarization [15,16]. (Note that since RNGH instability is the multimode amplitude instability [13], our numeric model utilizes slowly varying envelope approximation (SVEA) for the field amplitude and therefore it cannot reproduce spontaneous mode-jumps or slow phase instabilities).

What distinguishes RNGH instability from all other instabilities is that, when the regime of RNGH self-pulsations sets in, the medium polarization does not follow adiabatically the optical field in the cavity. However, at the initial stage when the CW single-mode lasing regime becomes unstable, there is only a small-amplitude perturbation of the optical field circulating in the cavity. In the most general case it has the form of an optical pulse. The interaction of the active medium with such a small-amplitude perturbation is still governed by adiabatic approximation [12] and does not necessarily give rise to a non-adiabatic behavior at a later stage. In order to define when the non-adiabatic behavior develops, we apply the pulse area theorem [19,21]. According to it, a perturbation in the form of an optical pulse is unstable and the pulse area grows if $\Omega_{Rabi} \cdot \tau_p > \pi$, where $\tau_p$ is the characteristic pulse width. Otherwise, the medium polarization just tracks the optical field behavior. The most general form of a small-amplitude perturbation that initially circulates in the cavity is such that its characteristic time is roughly a half of the cavity round-trip time $Ln_g/c$. Using also the fact that the maximum gain for instability is shifted from the initial (main) lasing mode by the Rabi frequency [$\Omega_{max} \approx \pm\Omega_{Rabi}$ in Fig. 2(a)], we recast the pulse area theorem into the form of instability condition at the maximum gain frequency $|\Omega_{max}|/2\pi \geq c/2Ln_g$. Indeed, in Fig. 2(a), the first case where the gain maximum is on the first non-lasing cavity mode occurs at $p\sim2.5$, when the numerical simulations do reveal the RNGH self-pulsations [see Fig. 2(b)]. The dashed vertical line in Fig. 2(b) shows the exact location of the second threshold obtained from the unstable pulse area condition. Because of the large frequency separation between the cavity modes, the second threshold raises considerably in short-cavity devices. At the same time, the threshold condition originating from the pulse area theorem has a minor impact on QCLs with long cavities, where mode separation frequency $c/2Ln_g$ is small. Thus for QCLs with cavity lengths of 3-4 mm, our linear stability analysis predicts that the second threshold is at a few percent above the lasing threshold [10], in agreement with available experimental data [4,7,8]. In addition, our numerical simulations based on TW rate equations model confirm the RNGH multimode operation at small excess above the lasing threshold [10].

At first glance, it is quite straightforward to solve the eigenproblem of the matrix from Eq. (2) and to find an analytic expression for the pump rate $p_{th2}$ at which one fulfills *the second threshold condition*:

$$\left|\Omega_{max}^{(th2)}\right|/2\pi = c/2Ln_g \qquad (6)$$

for the spectrum of the positive Lyapunov exponent. However the characteristic equation for the Lyapunov exponent is of the fourth order. Therefore we use a perturbation-theory approach. Since the matrix in Eq. (2) is not Hermitian, its eigenvectors are not orthogonal. For that reason we use the second-order bi-orthogonal perturbation theory [22,23] and split the linear stability matrix into the zero-order approximation $M^{(0)}$ and perturbation $M^{(1)}$ matrices, as indicated in Eq. (2). An example of the eigenvalue spectra of the initial matrix $M$ (solid black curves) and its zero-order approximation $M^{(0)}$ (red dashed curves) for a QCL pumped at p=1.5 times above the lasing threshold is depicted in Fig.3.

The eigenvalue with the largest real part $\Lambda_{max}$ is associated with the upper 2×2 block in $M^{(0)}$, which is independent of $\Omega_{Rabi}$. In the zero order approximation, the increment $Re(\Lambda_{max}^{(0)})$ just barely increases with the pump rate $p$ due to the SHB effect accounted for in the parameter $v_0$:

$$\Lambda_{max}^{(0)} = -\frac{1}{2T_{2,eff}} - \frac{1}{4\tau} - i\frac{\Omega}{2} + \frac{1}{2}\sqrt{\left(\frac{1}{T_{2,eff}} + \frac{1}{2\tau} + i\Omega\right)^2 - \frac{4i\Omega}{T_{2,eff}} + 4\frac{v_0-1}{T_{2,eff}\tau}}, \qquad (7)$$

where $\tau = n_g/cl_0$ is the photon lifetime in the cavity. Note that $\Lambda_{max}^{(0)} = 0$ at the lasing threshold ($p$=1) and zero detuning from the main lasing mode ($\Omega = 0$). The second eigenvalue originating from this 2×2 block has the largest negative real part, of $\sim \Lambda_{max}^{(0)} - 1/T_{2,eff} - 1/2\tau$ (see Fig. 3). In contrast to this behavior [See Eq.(7)], the two eigenvalues originating from the lower 2×2 block in $M^{(0)}$ are independent of the spectral detuning $\Omega$ but exhibit splitting that varies with $\Omega_{Rabi}$. At the lasing threshold ($p$=1) the splitting is of $1/T_{2,g} - 1/T_g$ and decreases with increasing $\Omega_{Rabi}$ (Note that $\Omega_{Rabi}^2 \ll T_{2,eff}^{-2}, T_{2,g}^{-2}$).

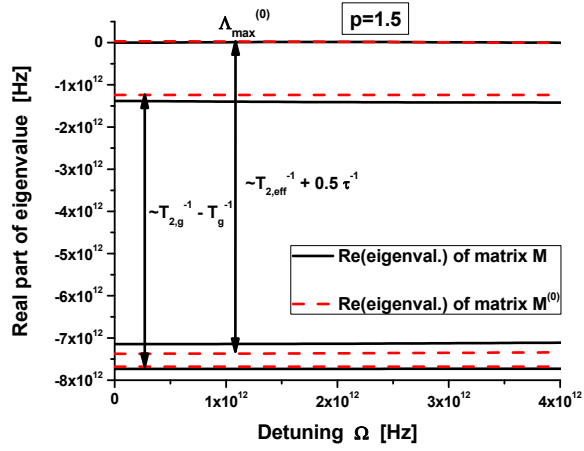

Fig. 3. Real part of eigenvalues of matrices M and $M^{(0)}$ from Eq. (2) for QCL with the cavity length L = 4 mm at pump rate $p$=1.5 times the lasing threshold. The red dashed curves show the real part of the eigenvalues of matrix $M^{(0)}$ and the black solid curves represent real part of eigenvalues of matrix M. The parameters used in simulations are listed in the column "QCL 1" in Table 1.

It can be seen from the example in Fig. 3 plotted for p= 1.5, that the eigenvalues of the matrices M and $M^{(0)}$ are close to each other as compare to the splitting. The perturbation matrix $M^{(1)}$ has no diagonal elements. Therefore the first-order correction terms to the eigenvalues of $M^{(0)}$ vanish. Hence we apply the second order of bi-orthogonal perturbative expansion in Eq. (2)

$$\Lambda_i = \Lambda_i^{(0)} + \frac{\langle U_i^{(0)}|M^{(1)}|V_i^{(0)}\rangle}{\langle U_i^{(0)}|V_i^{(0)}\rangle} + \sum_{\substack{m=1,\\m\neq i}}^{4} \frac{\langle U_m^{(0)}|M^{(1)}|V_i^{(0)}\rangle \langle U_i^{(0)}|M^{(1)}|V_m^{(0)}\rangle}{(\Lambda_i^{(0)} - \Lambda_m^{(0)})\langle U_m^{(0)}|V_m^{(0)}\rangle \langle U_i^{(0)}|V_i^{(0)}\rangle}, \qquad (8)$$

which provides a realistic approximation for the spectrum of RNGH instability gain Re($\Lambda_{max}$) in the vicinity of its maxima at $\Omega \sim \pm\Omega_{Rabi}$ (see Fig. 4, which we will discuss below). In Eq. (8), the diagonal term due to the first-order correction vanishes while the second-order correction terms are of the order of $\Omega_{Rabi}^2 T_{2,eff}$, $\Omega_{Rabi}^2 T_{2,g}$, which is much smaller than the initial spectral splitting indicated in Fig. 3. This justifies our decomposition onto the zero-order approximation $M^{(0)}$ and perturbation $M^{(1)}$ matrices in Eq.(2). In (8), the index $i$ varies from 1 to 4, $\Lambda_i$ denotes the eigenvalues, $V_i^{(0)}$ and $U_i^{(0)}$ are the concomitant partners, that is, these are the eigenvectors associated with the eigenvalues $\Lambda_i^{(0)}$ and $\Lambda_i^{(0)*}$ of the matrix $M^{(0)}$ and its adjoint matrix $M^{(0)+}$ respectively. The vectors $V_i^{(0)}$ and $U_i^{(0)}$ are used as the biorthogonal basis in the expansion (8). After some algebra, we obtain the following approximate expression for the Lyapunov exponent with the largest real part:

$$\mathrm{Re}(\Lambda_{max}) = -\frac{1}{2\tau} + \frac{C_0(p)}{\Omega^2 + 1/T_{2,eff}^2} + \frac{C_1(p)}{\Omega^2 + A(p)^2} + \frac{C_2(p)}{\Omega^2 + 1/T_{2,eff}^2} + \frac{C_3(p)}{(\Omega^2 + 1/T_{2,eff}^2)^2} \qquad (9)$$

where coefficients $C_i(p)$ and $A(p)$ are independent of the offset frequency $\Omega$ (they are shown in the Appendix B). The first two terms in Eq. (9) originate from $\Lambda_{max}^{(0)}$ in Eq. (7). The last three terms are due to the second order of perturbative expansion. For all cases considered here, $\Omega_{Rabi}, A \ll 1/T_{2,eff}$ so as the zero order terms and the two last terms in Eq. (9) vary slowly in the vicinity of the RNGH gain maxima at $\Omega \sim \pm \Omega_{Rabi}$. Furthermore, since $C_0 \gg C_2, T_{2,eff}^2 C_3$, the spectral locations of the gain maxima (at $\partial \mathrm{Re}(\Lambda_{max})/\partial\Omega = 0$) are mostly defined by the second and third terms in Eq. (9), yielding

$$\Omega_{max}^2 \approx T_{2,eff}^{-2} \sqrt{-C_1/C_0} - A^2 \ . \tag{10}$$

Using the expressions from Appendix B for parameters $A$ and $-C_1/C_0$, we finally obtain that the highest gain for multimode RNGH instability is at the following offset frequencies:

$$\Omega_{max}^2 \approx \Omega_{Rabi}^2 \sqrt{\frac{1}{2}\frac{T_{2,g}}{T_{2,eff}}\left(1+\frac{2}{\Omega_{Rabi}^2 T_g T_{2,g}}\right)} - \frac{1}{T_g^2} \tag{11}$$

This expression is valid when $\Omega_{Rabi}^2 \ll T_{2,eff}^{-2}$ (or $p - v_0 \ll T_1/T_{2,eff}$), that is true in all cases of practical interest considered here. The frequency $\Omega_{max}$ in Eq. (11) monotonically increases with the pump rate (e.g. see Fig. 4). Using Eq. (4), we find a reciprocal relationship to Eq. (11):

$$p - v(p) = \frac{T_1 T_{2,eff}}{T_g T_{2,g}}\left[\sqrt{1+2\frac{T_{2,g}T_{2,eff}}{T_g^2}(1+T_g^2\Omega_{max}^2)^2} - 1\right] \tag{12}$$

In Appendix C we obtain a general solution of Eq. (12) with respect to $p$. It appears that the instability increment (9) is negative and multimode instability is impossible at very small pump rates, when $p < p_{min}$, where

$$p_{min} = 1 + \frac{T_{2,eff}}{T_{2,g}} + \left(\frac{T_1 T_{2,eff}}{T_g T_{2,g}} - \frac{T_g^2}{T_{2,g}^2}\right)\left[\sqrt{1+2\frac{T_{2,g}T_{2,eff}}{T_g^2}} - 1\right] \approx 1 + \frac{T_{2,eff}^2}{T_g^2}\left[\frac{1}{2} + \frac{T_1}{T_g}\left(1 - \frac{T_{2,g}T_{2,eff}}{2T_g^2}\right)\right] \tag{13}$$

is the pump excess at which $\Omega_{max}^2 = 0$ in Eq. (12). For the pump rates above this value, the increment (9) is positive at the frequency (10) [see Fig. 2], which may lead to multimode instability. Finally to obtain the second threshold $p_{th2}$, we substitute in Eq. (12) our condition (6) for the $\Omega_{max}$ and we get

$$p_{th2} = 1 + \theta_{th2}\left(1 + \left[\frac{2T_1}{T_g} + \frac{T_{2,g}}{T_{2,eff}}\theta_{th2}\right]^{-1}\right), \quad \theta_{th2} = \frac{T_1 T_{2,eff}}{T_g T_{2,g}}\left[\sqrt{1+2\frac{T_{2,g}T_{2,eff}}{T_g^2}\left(1+\frac{c^2\pi^2 T_g^2}{L^2 n_g^2}\right)^2} - 1\right]. \tag{14}$$

The second threshold reduces towards $p_{min}$ [Eq. (13)] with increasing cavity length $L$. The meaning of $p_{min}$ corresponds to the notion of the second threshold in the original RNGH theory. Nevertheless by no means Eq. (13) can be directly compared with the second threshold for a unidirectional ring laser obtained in [5,6] (see also discussion in Section 3.3). However the shape of the multimode instability gain curve (9) can be assessed against the one in a unidirectional ring laser. For this purpose we provide an approximate expression for the maximum of increment (9) at the offset frequency (11), which we will use later:

$$\mathrm{Re}(\Lambda_{max})\big|_{\Omega=\Omega_{max}} = -\frac{1}{2\tau}\left(1 - \frac{3}{2}\frac{\Omega_{Rabi}^2 T_{2,eff}^2}{2}\right) + \frac{2v_0 - 1}{2\tau}\left[1 - \Omega_{max}^2 T_{2,eff}^2 - \frac{T_{2,eff}^2}{T_g^2}\right]^2 \tag{15}$$

## 3. Results and discussion

### 3.1 Validation of the analytic solution for the second threshold

In Fig. 4 we compare the instability gain curves obtained from the analytic expression (9) (red curves) and by solving numerically the eigenvalue problem of the linear stability matrix (2) (black curves). The results are plotted as the round-trip gain $2\mathrm{Re}(\Lambda)n_g L/c$ vs the offset frequency $\Omega/2\pi$. In this example we use a QCL with the cavity length of 4 mm and other parameters from the first column in Table 1. For the pump rate $p$ from 1 to 2.5 times above the lasing threshold and in the large spectral range, our analytical expression (9) is in reasonable agreement with the RNGH round-tip gain obtained by solving numerically for the eigenvalues of the matrix $M$ in Eq. (2). It is interesting to compare the instability gain curves for long –cavity QCL in Fig. 4 (the cavity mode separation is 11 GHz) with the ones obtained in Fig. 2(a) for a QCL with the cavity length of 100 µm (the intermodal frequency is 450 GHz). For concreteness we take the gain curves at $p$=2.5. Both gain curves are peaked at the same offset frequency of $\Omega_{max}/2\pi \sim 450$ GHz. Our Eq. (11) also shows that $\Omega_{max}$ is independent of the photon lifetime in the cavity.

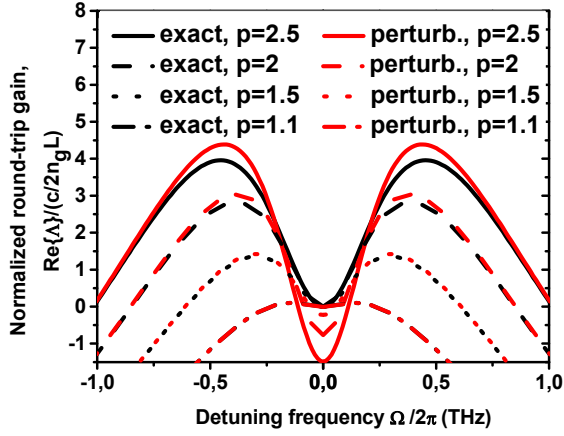

Fig. 4. The spectral behavior of the round-trip gain coefficient for multimode instabilities in QCL with the cavity length L = 4 mm (The cavity mode separation $c/2Ln_g$ =11 GHz) at different pump rates $p$ in the range from 1 to 2.5 times above the lasing threshold. The red curves show the gain spectra obtained from the analytical expression (9) and the black curves are obtained by solving numerically the eigenproblem of the matrix (2). The parameters used in simulations are listed in the column "QCL 1" in Table 1.

Figure 5 provides a comparison between the frequency $\Omega_{max}/2\pi$ obtained from the analytical expression (11) (red curve) and from the numerical solution of the eigenproblem of the matrix (2) (black curve). Our analytical expression for $\Omega_{max}$ is in reasonable agreement with the one obtained numerically. The $\Omega_{max}$ frequency approaches the Rabi oscillation frequency $\Omega_{Rabi}$ at high $p$ (blue curve). In order to illustrate the second threshold condition (6) we show a horizontal line at $\Omega/2\pi=c/2Ln_g$ for the cavity length of $L$=100 µm. Its crossing point with $\Omega_{max}/2\pi$ curve defines the second threshold, which is indicated by a vertical line. Our analytic approach provides a realistic approximation for the second threshold (the crossing point of the horizontal line with the red curve).

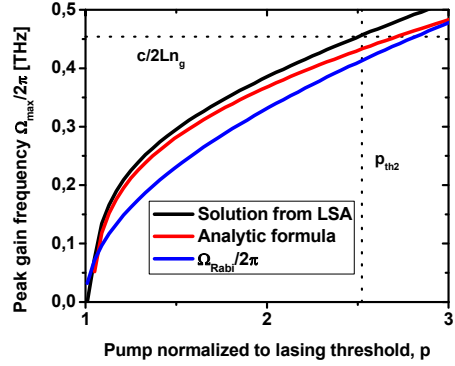

Fig. 5. Peak gain frequency of RNGH instability calculated numerically from our linear stability matrix (black curve) and from our analytic expression Eq. (10) (red curve) is plotted vs. pump normalized to lasing threshold $p$. It is shown in comparison with the behavior of Rabi oscillations $\Omega_{Rabi}/2\pi$ (blue curve). According to Eq. (6), the intersection between horizontal dotted line at $c/2Ln_g$ and the peak gain frequency defines the value of second threshold (vertical dotted line) in case of the 100 μm long cavity.

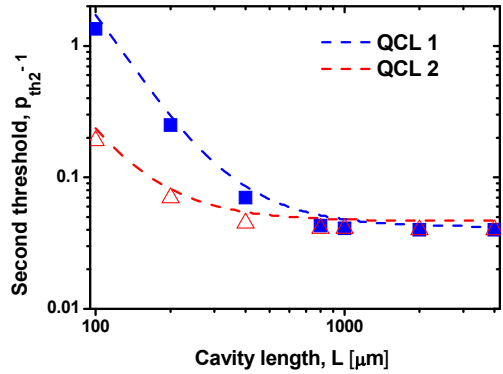

Fig. 6. Second threshold (represented as a relative pump excess above the lasing threshold $p_{th2}$-1) vs. cavity length. We compare $p_{th2}$ calculated from our Eq. (14) (curves) and calculated by numerical solving of the eigenproblem of the matrix (2) (squares and triangles). The parameters for QCL 1 and QCL 2 are listed in Table 1.

Figure 6 shows variation of the second threshold with the length of QCL cavity. (We plot the relative excess above the lasing threshold $p_{th2}$ -1 in logarithmic scale.) We consider two sets of QCL parameters from Table 1. In both cases, our analytical expression (14) (curves) shows good agreement with the numerical solutions for the eigenvalues of the linear stability matrix (2) (points). Note that the second threshold behavior predicted in Fig. 6 is in agreement with the results of numerical simulations based on the TW model [10] (see also Fig. 2) and with the numerous experimental reports indicating low second threshold in QCLs with the cavity length of 2-4 mm [4,7,8].

Our small signal analysis is based on a truncated system of coupled-mode equations [10] and our analytical expressions are valid in the limited range of the pump rates, when $p - \nu_0 \ll T_1/T_{2,eff}$. In the case of QCLs, we are practically limited to the range of $p<3$. For conventional QW LDs that have large gain recovery time $T_1$ (right column of Table 1) this condition implies that $p \ll 10^4$. However since we account only for the first nine coupled-mode equations in (1) and do not analyze the truncation error at high pump rates, our analysis cannot be justified for such high values of *p*. Nevertheless, it predicts a qualitatively reasonable behavior. More specifically, we formally obtain from Eq. (13) that $p_{min} \sim 1760$ for a QW LD with parameters shown in Table 1. Such second threshold is prohibitively high for experiment. Indeed, no experimental observation of RNGH instability has been reported for single-section QW LDs. These considerations also agree with the numerical simulations [10] indicating the onset of RNGH instability in single-section QW LDs for the pump rates as high as p~400 times above the lasing threshold.

### 3.2 The role of carrier population and carrier coherence gratings

The lowest second threshold [Eq. (13)] can be achieved in devices with very long cavities (see Fig. 6). Since in almost all practical cases $T_{2,g}T_{2,eff} \ll T_g^2$, Eq. (13) can be further simplified and represented in the following form

$$p_{min} \approx 1 + \frac{T_{2,eff}^2}{T_g^2}\left[\frac{1}{2} + \frac{T_1}{T_g}\left(1 - \frac{T_{2,g}T_{2,eff}}{2T_g^2}\right)\right]. \tag{16}$$

It can be seen that the coherence grating, whose relaxation time is $T_{2,g}$, does lower the second threshold but this effect is not as strong as a reduction due to the carrier population grating. The slower relaxation rate of the carrier population grating $1/T_g$, the lower the second threshold $p_{min}$. At first glance, the coherence grating just marginally enhances the impact of the SHB effect and therefore one may question whether the effect discussed here is indeed due to multimode RNGH instability. Moreover, the multimode instability behavior we discuss here is related to a build-up of macroscopic coherences in the medium [6], but the second threshold (16) increases when the decoherence rate $1/T_{2,eff}$ reduces. Note however that the second threshold for a *unidirectional* ring laser in the original RNGH theory [5,6] exhibits exactly the same counterintuitive behavior: *it increases with decreasing decoherence rate*.

If we exclude the effect of coherence grating ($T_{2,g} \to 0$) while simultaneously maintaining the effect of SHB on carrier distribution ($T_g, T_{2\_eff} \neq 0$), the system still reveals instabilities with the spectral shape similar to the one depicted in Fig. 2(a). In Gordon et al. [7], the following expression was obtained for the frequency at the maximum increment of the multimode instability caused by the SHB effect:

$$\Omega_{SHB}^{[7]} = \sqrt{\frac{1}{T_1}\sqrt{\frac{p-1}{3T_1T_2}}} \tag{17}$$

In Fig. 7 we plot this $\Omega_{SHB}^{[7]}$ frequency (green curve) as a function of the pump rate and using the characteristic time parameters from [7] (QCL2 in Table 1). We also show the frequency $\Omega_{max}$ for multimode RNGH instability (11) obtained in this case ($T_{2g} \neq 0$, solid blue curve) and the corresponding Rabi flopping frequency (4) (dashed blue curve). Because the frequency $\Omega_{SHB}^{[7]}$ is much smaller than $\Omega_{Rabi}$, the instability (17) has been attributed in [7] to a different kind of SHB-induced multimode instability. However, as mentioned in the Introduction, a set of inaccuracies was admitted in [7]. They are all listed in our recent work [10], a major consequence of which is that the coupled mode equations in [7] do not show a proper adiabatic transition to the case when the medium polarization varies slowly (the case of a usual FP cavity laser or a ring laser with *two counter propagating lasing modes*). Another consequence, which we discuss just below, is that the frequency of this instability is in fact very different from Eq. (17).

With our corrected system of coupled-mode Eqs. (2), in the case of $T_{2,g} \to 0$ (only SHB grating is present) we find that the effective threshold (5) due to the SHB effect is no longer bounded above by the value of $1 + T_{2,eff}/T_{2,g}$. Instead, it shows a linear growth

$$v_0^{(SHB)} = 1 + T_g(p-1)/(2T_1 + T_g), \tag{5'}$$

which is depicted in Fig. 1(a) by the dashed blue curve. As a result, both the Rabi frequency and the maximum increment frequency $\Omega_{max}$ (11) decrease. We obtain the following approximate expressions for these two frequencies when $T_{2,g} \to 0$:

$$\left(\Omega_{max}^{(SHB)}\right)^2 \approx \left(\Omega_{Rabi}^{(SHB)}\right)^2 \sqrt{\frac{1}{2}\frac{v_0^{(SHB)}}{(2v_0^{(SHB)}-1)(v_0^{(SHB)}-1)}} - \frac{1}{T_g^2}, \quad \Omega_{Rabi}^{(SHB)} = \sqrt{\frac{2(p-1)}{(2T_1+T_g)T_{2,eff}}}. \tag{11'}$$

These are also plotted in Fig. 7 (dashed and solid red curves).

The following observations can be made in Fig. 7: (i) With coherence grating ($T_{2,g} \neq 0$), the highest instability increment occurs for the modes at the offset frequency $\Omega_{max}$ [Eq. (11)] which is very close to the Rabi frequency from Eq. (4) (note an almost linear growth of $\Omega_{Rabi}^2$ in the scale of the Fig. 7). This behavior is typical for multimode RNGH instability [5,6,7,13]. (ii) Without coherence grating ($T_{2,g} \to 0$), the Rabi flopping frequency $\Omega_{Rabi}^{(SHB)}$ decreases and $\Omega_{max}^{(SHB)}$ [Eq. (11')] deviates from the behavior of the Rabi oscillation frequency $\Omega_{Rabi}^{(SHB)}$. However, at p=3, it is only by a factor of 1.15 smaller than $\Omega_{Rabi}^{(SHB)}$. (In [10] we calculate it numerically and find that the difference is even smaller.) Our frequency $\Omega_{max}^{(SHB)}$ is still much higher than the frequency $\Omega_{SHB}^{[7]}$ [Eq. (17)] obtained in [7] for multimode instabilities caused by SHB effect.

The didactic question whether or not the multimode instability at frequency $\Omega_{max}^{(SHB)}$ [Eq. (11')] can be associated with the RNGH behavior occurring without coherence grating we leave for a discussions outside of the scope of this paper. The opposite case when the coherence grating is taken into account is much more significant. Indeed due to diffusion processes, both SHB and coherence grating can be smoothed only simultaneously [see the definitions of the relaxation times $T_{2,g}$ and $T_g$ in Eq. (2)]. Most importantly, even though the effect of the coherence grating might be deemed as a small correction to the behavior caused by the SHB effect, the resulting frequency for the maximum instability increment $\Omega_{max}$ [Eq. (11)] nearly follows the behavior of the Rabi oscillations frequency $\Omega_{Rabi}$ [Eq. (4)], attesting for the RNGH-like behavior (see Fig. 7).

In Section 2 we have mentioned the two submatrices (5×5 and 4×4 size) in the 9×9 block-diagonal matrix (1) describing the linear stability of our laser system. We have explained that due to the SHB effect and induced coherence grating, only the 4×4 matrix may lead to RNGH-like behavior in the range of pump rates considered here (see also [10]). In the presence of coherence grating, the 5×5 matrix is stable up to very high pump rates (up to $p \sim 50$ times above the lasing threshold in QCL 1 and up to p~40 for QCL 2 in Table 1). That is, the 5×5 matrix is stable over the entire range 9<$p$<15 [10] predicted by the original RNGH theory. Only when $T_{2,g} \to 0$ the 5×5 matrix exhibits RNGH instability in the usual range, in parallel with the instability at $\Omega_{max}^{(SHB)}$ originating from the 4×4 matrix.

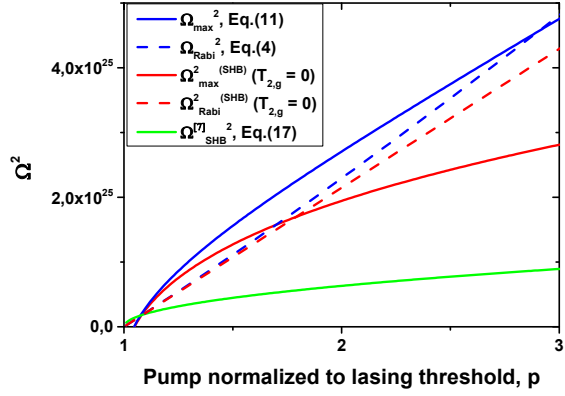

Fig. 7. Using parameters of QCL2 from Table 1, we plot squares of the following frequencies as a function of the pump rate: $\Omega_{max}$ frequency for multimode RNGH instability from Eq. (11) obtained for $T_{2,g} \neq 0$ (solid blue curve) and the corresponding Rabi frequency $\Omega_{Rabi}$ (4) (dashed blue curve), the frequency $\Omega_{max}^{(SHB)}$ for multimode instability occurring without coherence grating (when $T_{2,g} \to 0$, solid red curve) and its corresponding Rabi frequency $\Omega_{Rabi}^{(SHB)}$ (dashed red curve) as well as the frequency $\Omega_{SHB}^{[7]}$ for this instability calculated in [7] (green curve).

### 3.3 Link to the original RNGH theory

In the limit of $T_g, T_{2,g} \to 0$, that is when we exclude *both* carrier (SHB) and coherence gratings, our 9×9 linear stability matrix reduces to a model for an unidirectional ring laser, like the one considered in [5,6]. In this case, the submatrix corresponding to our 4×4 matrix (2) is always stable [7], while the other one originating from the 5×5 block exhibits the RNGH instability for $p\sim10$ (for QCLs in Table 1) or $p\sim9$ (for LD), in agreement with the original RNGH theory [5,6,13]. Since our estimate for the second threshold (14) originates from a different matrix block, by no means it may show a limiting transition to the original RNGH case of a unidirectional ring laser and recover the magic value for the second threshold of p=9. However the behavior of the multimode instability gain curve (9) can be assessed against the one in a ring laser. Below we provide approximate expressions for the maximum of instability increment (15) and compare it with the one obtained for RNGH instability in a unidirectional ring laser [13]. In both cases these maxima occur at respective frequencies $\Omega_{max}$ quite close to the Rabi frequency $\Omega_{Rabi}$.

First of all, we note that there is no effective threshold increase (5) in a unidirectional ring laser. Therefore to draw parallels between RNGH instability in a ring laser and the multimode instability in FP laser discussed here, we substitute $\nu_0=1$ in Eq. (15). Next, using that $\Omega_{Rabi} \ll 1/T_{2,eff}$, we find that the maximum increment for multimode instability in a FP cavity laser can be represented in the following form:

$$\mathrm{Re}(\Lambda_{max})\big|_{\Omega=\Omega_{max}} \approx \frac{3}{8\tau}\Omega_{Rabi}^2 T_{2,eff}^2 - \frac{1}{\tau}\Omega_{max}^2 T_{2,eff}^2 , \qquad (18)$$

where the frequencies $\Omega_{Rabi}$ and $\Omega_{max}$ are related by Eq. (11).

For the RNGH instability gain curve in a unidirectional ring laser, we use an expression from [13], Lugiato et al. (we refer to the Eq. (22.50) in there). Once again, using that $\Omega_{Rabi} \ll 1/T_2$, we recast it into a simple form and obtain its peak value of

$$\text{Re}(\Lambda_{\max}^{(RNGH)})\Big|_{\Omega=\Omega_{\max}^{(RNGH)}} \approx \frac{3}{2\tau}\Omega_{Rabi}^2 T_2^2 - \frac{1}{\tau}\left(\Omega_{\max}^{(RNGH)}\right)^2 T_2^2, \qquad (19)$$

Where $\Omega_{Rabi}$ is given by Eq. (4) [with $v_0$=1]. However the maximum increment is reached at a different frequency $\Omega_{\max}^{(RNGH)}$ [13] (see also Eq. (19) in [7]):

$$\Omega_{\max}^{(RNGH)} \approx \Omega_{Rabi}\sqrt[4]{2}, \quad \Omega_{Rabi} = \sqrt{\frac{p-1}{T_1 T_2}} \qquad (20)$$

Note that the two expressions, the Eq. (19) for RNGH instability in a unidirectional ring laser and our Eq. (18) for a FP cavity laser are surprisingly similar, indicating the same RNGH-like behavior in both cases.

## 4. Conclusion

This paper provides an analytical expression for the second threshold (RNGH instability threshold) in a quantum cascade laser with FP cavity. This simple analytical expression is a versatile tool that can easily be applied in practical situations requiring analysis of QCL dynamic behavior and estimation of its second threshold. We show that the second threshold in QCLs is low due to the induced carrier (SHB) and coherence gratings and it is much lower than in conventional bipolar semiconductor LDs, in agreement with various experiments reported in literature. The future work will be to take the inhomogeneous broadening of the gain curve and it shape into account, the feature that directly affects the observation of the QCL's RNGH instability in experiment.

## Appendix A

The slowly varying envelope approximation for the coupled-mode rate equations model reads (the details can be found in Ref. [10])

for the field amplitudes:
$$\frac{n_g}{c}\partial_t E_\pm = \mp\partial_z E_\pm - i\frac{N\mu\Gamma\omega}{cn_g\varepsilon_0}\eta_\pm - \frac{1}{2}l_0 E_\pm, \qquad (21)$$

for the medium polarization:
$$\partial_t \eta_\pm = \frac{i\mu}{2\hbar}\left(\Delta_0 E_\pm + \Delta_2^\mp E_\mp\right) - \frac{\eta_\pm}{T_2} - k^2 D\eta_\pm, \qquad (22)$$

for the coherence grating amplitude:
$$\partial_t \eta_{\pm\pm} = \frac{i\mu}{2\hbar}E_\pm \Delta_2^\mp - \frac{\eta_{\pm\pm}}{T_2} - 9Dk^2 \eta_{\pm\pm}, \qquad (23)$$

for the carrier population density:
$$\partial_t \Delta_0 = \frac{\Delta_{pump} - \Delta_0}{T_1} + \frac{i\mu}{\hbar}\left(E_+^* \eta_+ + E_-^* \eta_- - c.c\right), \qquad (24)$$

and population grating:
$$\partial_t \Delta_2^\pm = \frac{i\mu}{\hbar}\left(E_\pm^* \eta_\mp - E_\mp \eta_\pm^* - E_\pm \eta_{\pm\pm}^* + E_\mp^* \eta_{\mp\mp}\right) - \frac{\Delta_2^\pm}{T_1} - 4k^2 D\Delta_2^\pm. \qquad (25)$$

Here, $\omega$ and $\mu$ denote the resonant frequency and the dipole matrix element of the optical gain transition. The longitudinal and transverse relaxation times are labeled as $T_1$ and $T_2$; $\Delta_{pump}$ is the steady-state population inversion, and $D$ is the diffusion coefficient. $E$, $N$, and $\Gamma$ stand for the optical field, the number of two-level systems per unit volume and the overlap factor between the optical mode and the active region, $n_g$ is the background refractive index. The two optical carrier waves propagating at frequency $\omega$ and wavenumbers $\pm k$ define the initial lasing mode in the cavity. The spatial grating of the carrier population is taken into account through the terms $\Delta_2^\pm$, while the spatial amplitudes of the induced macroscopic polarization grating are labeled as $\eta_{++}$ and $\eta_{--}$. The cavity loss coefficient $l_0$ comprises the intrinsic material losses $l_0$' and the output coupling losses. By solving Eqs. (21)-(25) for the single mode CW lasing regime and introducing small variations around the steady state values of variables we arrive to Eq.(1).

**Appendix B**

The coefficients $C_0, C_1, C_2$ and $C_3$ in Eq. (9) read:

$$C_0 = -\frac{\rho}{\tau T_{2,eff}} \tag{26}$$

$$C_1 = \frac{1}{\tau}\frac{E}{A^2 - T_{2,eff}^{-2}}\left(AE\rho + \frac{A\alpha}{T_{2,eff}} + \frac{2E\rho}{T_{2,eff}} + A^2\alpha\right) + \frac{2}{\tau}\frac{E^2\rho(A+T_{2,eff}^{-1})}{(A^2 - T_{2,eff}^{-2})^2 T_{2,eff}^2}, \tag{27}$$

$$C_2 = -\frac{1}{\tau}\frac{E}{A^2 - T_{2,eff}^{-2}}\left(AE\rho + \frac{A\alpha}{T_{2,eff}} + \frac{2E\rho}{T_{2,eff}} + \frac{\alpha}{T_{2,eff}^2}\right) - \frac{2}{\tau}\frac{E^2\rho(A+T_{2,eff}^{-1})}{(A^2 - T_{2,eff}^{-2})^2 T_{2,eff}^2} \tag{28}$$

$$C_3 = \frac{2}{\tau}\frac{E^2\rho(A+T_{2,eff}^{-1})}{(A^2 - T_{2,eff}^{-2})T_{2,eff}^2} \tag{29}$$

where we have introduced the following parameters:

$$\rho = -\frac{2\nu_0 - 1}{2T_{2,eff}}, \quad \alpha = \frac{E}{2}\left[1 + (\nu_0 - 1)\frac{T_{2,g}}{T_{2,eff}}\right], \quad A = \frac{1}{2T_g} + \frac{1}{2T_{2,g}} - \frac{1}{2}\sqrt{\left(\frac{1}{T_{2,g}} - \frac{1}{T_g}\right)^2 - 4E^2} \tag{30}$$

This solution is valid when $\Omega_{Rabi} < T_{2,eff}^{-1}$ ($T_{2,eff}^{-1} \leq T_{2,g}^{-1}$ due to the diffusion terms). In the range of pump rates around the second threshold, $\Omega_{Rabi}^2 \ll T_{2,eff}^{-2}$ (which assumes $p - \nu_0 \ll T_1/T_{2,eff}$) and hence further simplification of Eqs. (26-30) is possible, yielding, with accuracy to the correction terms of relative order $\Omega_{Rabi}^4 T_{2,eff}^2$ in Eq. (9),

$$C_1 \approx -\frac{1}{2\tau}\frac{T_g T_{2,g} E^2}{(T_g - T_{2,eff})^2}\left[\nu_0\left(1 + 2\frac{T_{2,eff}}{T_{2,g}}\right) - 1\right]\cdot\left(1 + \frac{E^2 T_g^3 T_{2,g}}{(T_g - T_{2,g})(T_g - T_{2,eff})}\right) \tag{31}$$

$$C_2 \approx \frac{1}{2\tau}\frac{T_g^2 T_{2,g} E^2}{T_{2,eff}(T_g - T_{2,eff})^2}\left[\nu_0\left(1 + \frac{T_{2,eff}}{T_g}\right) - 1 + \frac{T_{2,eff}}{T_{2,g}}\right] \tag{32}$$

$$C_3 \approx \frac{1}{\tau}\frac{E^2 T_g (2\nu_0 - 1)}{T_{2,eff}^2(T_g - T_{2,eff})}, \quad A \approx \frac{1}{T_g} + \frac{T_g T_{2,g} E^2}{T_g - T_{2,g}} \approx \frac{1}{T_g} + \frac{1}{2}T_{2,g}\Omega_{Rabi}^2 \tag{33}$$

It follows that

$$-\frac{C_1}{C_0} \approx T_{2,g}^2 T_{2,eff}^2 \Omega_{Rabi}^4 \frac{1}{4}\frac{\nu_0(1 + 2T_{2,eff}T_{2,g}^{-1}) - 1}{2\nu_0 - 1}\cdot\left(1 + \frac{2}{\Omega_{Rabi}^2 T_g T_{2,g}}\right). \tag{34}$$

Therefore $-C_1/C_0 \ll \frac{T_{2,g}}{T_g} \ll 1$ for the pump rates considered here.

**Appendix C**

In this appendix we solve an Eq. of the form $p - \nu(p) = \theta$ [see Eq. (12)] with respect to the pump rate $p$. Using expression for $\nu_0$ from Eq. (5), after rearrangement of terms and squaring, we obtain:

$$\left(\frac{2\theta - p + \phi}{2}\right)^2 = \left(\frac{p+\phi}{2}\right)^2 - p\left(\phi - \frac{2T_1 T_{2,eff}}{T_g T_{2,g}}\right)^2 - \frac{2T_1 T_{2,eff}}{T_g T_{2,g}} \tag{35}$$

Where we have introduced a new parameter

$$\varphi = 1 + \frac{T_{2,eff}}{T_{2,g}} + \frac{2T_1 T_{2,eff}}{T_g T_{2,g}} \tag{36}$$

Eq. (35) is of the first order with respect to variable p. Its solution reads

$$p = 1 + \theta\left(1 + \left[\frac{2T_1}{T_g} + \frac{T_{2,g}}{T_{2,eff}}\theta\right]^{-1}\right) \tag{37}$$

## Funding

Swiss National Science Foundation (project FASTIQ), Ministry of Education, Science and Technological Development (Republic of Serbia), ev. no. III 45010, COST ACTIONs BM1205 and MP1204, European Union's Horizon 2020 research and innovation programme under the grant agreement No 686731 (SUPERTWIN).

## Acknowledgments

DBo is grateful to Vladimir Kocharovsky and Ekaterina Kocharovskaya for valuable discussions.